\documentstyle[aps,prl,graphicx]{revtex}
\begin{document}

\title{\bf{Particle scattering in loop quantum
gravity}\\[.5mm]} \author{Leonardo Modesto and Carlo Rovelli\\ 
\em Centre de Physique Th\'eorique de Luminy, Universit\'e de la
M\'editerran\'ee, F-13288 Marseille, EU}
\date{\small \today}

\maketitle

\begin{abstract}

\noindent We devise a technique for defining and computing $n$-point
functions in the context of a background-independent gravitational
quantum field theory.  We construct a tentative implementation of this
technique in a perturbatively-finite loop/spinfoam model.

\end{abstract}

\vspace{.5cm}

\noindent The lack of a general technique for computing particle
scattering amplitudes is a seriously missing ingredient in
nonperturbative quantum gravity \cite{book,thomas}.  Various problems
can be traced to this absence: the difficulty of deriving the low
energy limit of a theory; of comparing alternative theories, such as
alternative versions of the hamiltonian operator in loop quantum
gravity (LQG) or different spinfoam models; or comparing the
predictions of a theory with those of perturbative approaches to
quantum gravity, such as perturbative string theory.  Here we explore
one possibility for defining a general formalism aimed at computing
scattering amplitudes.  We outline a calculation strategy, which can
certainly be improved.  Our interest is not in a particular theory,
but rather in a general technology to be used for analyzing different
models.  For concreteness, we implement this strategy in the context
of a specific model, and present a well-defined and perturbatively
finite expression, which, under substantial assumptions and
approximations, might be used as a general covariant $n$-point
function.

In conventional QFT, we can derive all scattering amplitudes from the
$n$-point functions
\begin{equation}
W(x_{1}, \ldots, x_{n})=Z^{-1}\int D\phi\ \phi(x_{1})\ldots
\phi(x_{n}) \ e^{-iS[\phi]}
\label{Wn}
\end{equation}
where the $x_{i},\ i= 1, \ldots, n$ are points of the background
spacetime, $\phi$ is the quantum field, $S[\phi]$ its action and $Z$
is the integral of the sole exponential of the action.  Alternatively,
the $n$-point functions can be derived from their Euclidean
continuations, defined by dropping the $i$ factor in the above
expression.  The integral (\ref{Wn}) is well-defined in perturbation
theory or as a limit of a lattice regularization, under appropriate
renormalization.  A well-known difficulty of background independent
quantum field theory is given by the fact that if we assume (\ref{Wn})
to be well-defined with general-covariant measure and action, then the
$n$-point function is easily shown to be constant in spacetime (see
for instance \cite{TW}).  This is the difficulty we address here.

Consider a spacetime region $R$ such that the points $x_{i}$ lie on
its 3d boundary $\Sigma$.  Call $\varphi$ the restriction of the field
$\phi$ to $\Sigma$.  Then (\ref{Wn}) can be written in the form
\begin{equation}
W(x_{1}, \ldots, x_{n})=Z^{-1}\int D\varphi\ \varphi(x_{1})\ldots
\varphi(x_{n})\ W[\varphi,\overline \Sigma] \ W[\varphi,\Sigma]
\label{WW}
\end{equation}
where (in the euclidean case)
\begin{equation}
 W[\varphi, \Sigma]=\int_{\phi|_{{}_{\Sigma}}=\varphi} 
 D\phi_{R}\  e^{-S_{R}[\phi_{R}]}
\label{W}
\end{equation}
is the integral over the fields in $R$ bounded by the 3d field
$\varphi$ on $\Sigma$, and $S_{R}$ is the restriction of the action to
$R$; while $W[\varphi, \overline\Sigma]$ is the analogous quantity
defined on the complement $\overline R$ of $R$ in spacetime.  The
field boundary propagator (\ref{W}) has been considered in
\cite{oeckl} and \cite{book} and studied in \cite{cr}.  Assume that we
are dealing with an interacting theory, approximated by a free
(gaussian) theory $S^{(0)}[\phi]$ in some regime, and that, within a
certain approximation of the amplitude (\ref{Wn}), the interaction
term in the action can be restricted to $R$.  Then we can replace
$W[\varphi,\overline R]$ with its free theory equivalent
\begin{equation}
 W_{0}[\varphi, \overline \Sigma] \ =\ 
\int_{\phi|_{{}_{\Sigma}}=\varphi} 
 D\phi_{\bar R}\ e^{-S^{(0)}_{\bar R}[\phi]}\ \equiv\
 \Psi_{\Sigma}[\varphi].
\label{W2}
\end{equation}
This integral is gaussian and can be performed, giving a gaussian
``boundary state" $\Psi_{\Sigma}[\varphi]$, determined by appropriate
boundary conditions for the field at infinity.  For instance, if we
take $R$ to be defined by $t>0$, then $\Psi_{t=0}[\varphi]$ is the
vacuum state in the functional Schr\"odinger representation.  In
general, we expect the boundary state to be given by some gaussian
functional of the boundary field $\varphi$ on $\Sigma$.

Consider a diffeomorphism invariant theory including the gravitational
field.  Assume that the equations above hold, in some appropriate
sense.  The field $\phi$ represents the gravitational field, as well
as any eventual matter field, and we assume action and measure to be
diffeomorphism invariant.  Two important facts follow \cite{cdort}. 
First, because of diffeomorphism invariance the boundary propagator
$W[\varphi,\Sigma]$ is independent from (local deformations of) the
surface $\Sigma$.  Thus in gravity the left hand side of (\ref{W})
reads $W[\varphi]$.  Second, the geometry of the boundary surface
$\Sigma$ is not determined by a background geometry (there isn't any),
but rather by the boundary gravitational field $\varphi$ itself.

We can obtain an indication on the possible forms of the boundary
state in gravity from the free quantum theory of non-interacting
gravitons on Minkowski space.  If we take $R$ to be $t>0$, for
instance, then $\Psi_{t=0}[\varphi]$ must be approximated by the
well-known Schr\"odinger vacuum wave functional of linearized 
gravity. 
This is a gaussian state picked around a classical geometry: the flat
geometry of the $t=0$ surface in Minkowski space.  In the case of a
compact $R$, it is then reasonable to consider a gaussian boundary
state $\Psi_{q}[\varphi]$ picked around \emph{some} 3-geometry $q$ of
the boundary surface $\Sigma$.  Thus, we may expect an expression of
the form
\begin{equation}
W(x_{1}, \ldots, x_{n};\ q)=Z^{-1}\int D\varphi\ 
\varphi(x_{1})\ldots
\varphi(x_{n})\ \Psi_{q}[\varphi]\ W[\varphi] 
\label{Wn4}
\end{equation}
to approximate (\ref{Wn}) when the interaction term can be neglected
outside $R$.  For this equation to be significant, we have to fix the
meaning of the coordinates $x_{i}$, since the rest of the expression
is generally covariant.  There is an obvious choice: the points
$x_{i}$ can be defined with respect to the geometry $q$.  For
instance, if $n=4$, $t^0_{1}=t^0_{2}=0$ and $t^0_{3}=t^0_{4}=T$, (we
use $x=(t,\vec x)$) we can take $q$ to be the geometry of a
rectangular box of height $T$ and side $L$ and interpret $\vec x_{i}$
as proper distances from the boundaries of the box.  In other words,
\emph{the localization of the arguments of the $n$-point function can
be defined with respect to geometry over which the boundary state is
picked}.  Notice that $x_{i}$ in (\ref{Wn4}) are then \emph{metric}
coordinates: they refer to gravitational field values.  They are not
anymore general-covariant coordinates as in (\ref{Wn}).  In this
manner, we can give meaning to $n$-point functions in a background
independent context.

Physically, we can interpret $R$ as a finite spacetime region where a
scattering experiment is performed.  The quantities $x_{i}$ are then
relative distances and relative proper time separations, measured
along the boundary of this region, and determined by (the mean value
of) the gravitational field (hence the geometry) on this boundary. 
This is precisely the correct general-relativistic description of the
position measurements in a realistic scattering experiment.

In order to give (\ref{Wn4}) a fully well-defined meaning, and compute
$n$-point functions concretely, we need four ingredients: (i) A proper
definition of the space of the 3d fields $\varphi$ integrated over,
and a well-posed definition of the integration measure.  (ii) An
explicit expression for the boundary propagator $W[\varphi]$.  (iii)
An explicit expression for the boundary state $\Psi_{q}[\varphi]$. 
(iv) A definition of the field operator $\varphi(x)$.  In the
following we analyze the status of these four ingredients in the loop
and spinfoam approach to quantum gravity.  We consider
for simplicity pure gravity without matter.

(i) In quantum theory, the boundary values of Feynman integrals can be
taken to be the classical dynamical variables only if the
corresponding operators have continuum spectrum.  If the spectrum is
discrete, the boundary values are the quantum numbers that label a
basis of eigenstates (see \cite{book}).  In our case, the boundary
field $\varphi$ represents the metric of a 3d surface.  Let us assume
here the results of LQG that the 3d metric is
quantized \cite{book}.  Therefore we must replace the continuum
gravitational field variable $\varphi$ with the quantum numbers
labelling a basis that diagonalizes some metric degrees of freedom. 
These can be taken to be (abstract) spin networks $s$, or $s$-knots. 
An $s$-knot is here an equivalence class under (extended
\cite{winston}) diffeomorphisms of embedded spin networks $S$.  An
embedded spin network is a graph immersed in space, labeled with spins
and intertwiners.  The $s$-knots are discrete \cite{winston}.  Thus,
we rewrite (\ref{Wn4}) in the form 
\begin{equation} 
W(x_{1}, \ldots
x_{n};\ g)=Z^{-1}\!  \sum_{s} c(s)\ \varphi(x_{1})\ldots
\varphi(x_{n}) \Psi_{q}[s]\ W[s].
\label{W4s}
\end{equation}
where the meaning of $\varphi(x)$ will be specified later on.  The
discrete measure $c(s)$ on the space of the $s$-knots is defined by
the projection of the scalar product of the space of the embedded spin
networks ($c(s)$=1) except for discrete symmetries of $s$.)  For
simplicity, and in order to match with the spinfoam formalism that we
use below, we restrict here the space of the spin networks to the
four-valent ones and we identify spin networks with the same graph,
spins and intertwiners (i.e., we ignore knotting and linking).

(ii) The boundary propagator $W[s]$ is a now a function of a boundary
spin network.  A natural possibility is to identify it with the
boundary propagator $W[s]$ defined by the spin foam models
\cite{Perez}.  For concreteness, let us choose here the model
defined by the $SO(4)/SO(3)$ group field theory \cite{PR}, which gives
a perturbation expansion finite at all orders \cite{finiteness}.  This
is the model denoted $GFT/C$ in \cite{book}.  The amplitude of a spin
network $s$ is given in this model by 
\begin{equation}
 W[s]=
 \int D\Phi\ f_{s}[\Phi]\ e^{-\int \Phi^2-\lambda \int \Phi^5}.
\label{GFT}
\end{equation}
Here $\Phi$ is a function on $[SO(4)]^4$ and the precise meaning of
the (symbolic) integrals in the exponent is detailed in \cite{Perez}
and \cite{book}.  The quantity $f_{s}[\Phi]$ is a polynomial in the
field $\Phi$, determined by $s$.  It is defined by
picking one factor
\begin{equation} 
\Phi^i_{\alpha_{1}\ldots\alpha_{4}} =
\int dg_1\ldots dg_4\ \Phi(g_{1},\ldots,g_{4})\ 
R^{(j_1)}_{\alpha_{1}}{}^{\beta_1}(g_{{1}}) \ldots 
R^{(j_4)}_{\alpha_4}{}^{\beta_{{4}}}(g_{{4}}) \ 
v^{i}_{\beta_{{1}}\ldots\beta_4}
\end{equation} 
per node of $s$, where $v^i$ is the intertwiner of the node and
$j_{1},\ldots, j_{4}$ the colors of the adjacent links, and
contracting the indices $\alpha_{i}$ according to the connectivity of
the graph of $s$.  The expression (\ref{GFT}) is well-defined and
finite order by order in $\lambda$.  (The rigorous proof of this
statement is complete up to certain degenerate graphs \cite{Perez}.) 
The explicit computation of $W[s]$ is entirely combinatorial and can
be performed in terms of combinations of $nJ$ Wigner symbols
\cite{book}.  For completeness, recall that the reason for the
definition (\ref{GFT}) is that the expansion of $W[s]$ in $\lambda$
can be written as a sum over spinfoams bounded by the spin network $s$
\begin{equation}
 W[s]=\sum_{\partial\sigma=s} A(\sigma),
 \label{sf}
\end{equation}
where the spinfoam amplitude $A(\sigma)$ is the Barrett-Crane 
discretization of
the exponential of the Einstein-Hilbert action of the discrete
four-geometry defined by the spinfoam $\sigma$.  Therefore the
definition (\ref{GFT}) of $W[s]$ can be interpreted as a (background
independent) discretization of the functional integral (\ref{W}).

(iii) An expression for the boundary state $\Psi_{q}[s]$ can be
obtained from the analysis of the coherent states in LQG
\cite{coherent,coherent2,florian}.  For concreteness, let us pick here
Conrady's definition of a coherent state \cite{florian}.  Other more
refined expression could be used instead.  Conrady has defined a state
$\Psi_0[S]$ that describes the Minkoski vacuum as a function of
embedded spin networks $S$, under certain approximations and
assumptions.  This function has the property of being picked on spin
networks that are ``weaves", namely that approximate a flat metric $q$
when averaged over regions large compared to the Planck scale
\cite{weave}.  This vacuum state can be written as follows.  Pick
cartesian coordinates $x^a, a=1,2,3$, on a 3d surface equipped with a
flat metric $q$ and with total volume $V$.  Fix a triangulation
$\mathcal{T}$ of lattice spacing $a$, small compared to the Planck
length $l_{p}$ in the metric $q$.  Restrict the attention to embedded
spin networks $S$ living on $\mathcal{T}$.  Define the form factor of
a spin network as
\begin{eqnarray}
F^{ab}_{S}(\vec{x}) & = &\frac{\pi l_P^4}{96a^3} \, 
\sum_{v\in S}\sum_{e,e'\in v}\int_0^1dt \int_0^1dt^{\prime} \,
       \dot{e}^a(t) \, \dot{e}^b(t^{\prime}) 
\, \delta(\vec{x} - \vec{x}_v),
\end{eqnarray}
where $v$ are the vertices of the spin network $S$; $\vec x_{v}$ their
position; $e:t\mapsto e^a(t)$ the edges; $e \in v$
indicates the edges $e$ adjacent to the vertex $v$; and $\dot
e^a=de^a/dt$.  Its Fourier transform is $
F^{ab}_{S}(\vec{k})  =  V^{-1} \int d^3 x \
\mbox{e}^{- \mbox{i} \vec{k} \cdot \vec{x}}
         \  F^{ab}_{S}(\vec{x}).$
Then
\begin{eqnarray}
\Psi_{0}[S] = {\mathcal{N}} \, \mbox{exp}\Big[-\frac{1}{4 \, l_p^2} 
\, 
\sum_{\vec{k}} |\vec{k}|\ 
    \big| F^{ab}_{S}(\vec{k}) \, 
j_e(j_e +1) - \sqrt{V} \delta^{ab} \, 
       \delta_{\vec{k},0} \big|^2 \Big], 
       \label{vacuum2}
\end{eqnarray}
where the momenta summed over are the discrete modes on the
triangulation; $j_{e}$ is the spin associated to the edge $e$; $
\mathcal{N}$ is a normalization factor.  To understand this
construction, notice that if we consider the gravitational field
associated to the spin network, (in the sense of the weaves)
$q_{S}^{ab}(\vec{x}) = F^{ab}_{S}(\vec{x}) \, j_e(j_e +1)$, then
$\Psi_0[S]=\Psi_0[q_S]$ where
\begin{eqnarray}
\Psi_0[q] & = &  \mbox{e}^{-\frac{1}{4 \hbar \kappa} \int 
d^3x\int d^3y \,
      [(\tilde{q}^{ab}(\vec{x}) - \delta^{ab}) \, W_{\Lambda}(\vec{x} 
- \vec{y})  \,
      (\tilde{q}^{ab}(\vec{y}) - \delta^{ab})]} 
      \label{vacuum1}
\end{eqnarray}
is the Sch\"odinger functional representation of the linearized vacuum
state.  $W_{\Lambda}(\vec{x} - \vec{y})$ is a lattice regularization
of the vacuum covariance.  We can extend this construction to a 3d
(euclidean) rectangular boundary $\Sigma$ simply taking the product of
the Conrady states associated to each of the eight faces forming
$\Sigma$.

We need to carry this result over to the diffeomorphism invariant
$s$-knot states.  Given an abstract spin network $s$ there will be in
general one embedded spin network $S(s)$ that maximizes the state
$\Psi_{0}[S]$.  We can then tentatively define
$\Psi_{0}[s]=\Psi_{0}[S(s)]$.  Notice that if $\Psi_{0}[S]$ is picked
on weaves, then the diffeomorphism invariant state $\Psi_{0}[s]$
defined is picked on the corresponding (``weavy") discrete
3-geometries.  The maximization condition can be interpreted as a
gauge choice, picking the coordinate system in which the 3-geometry is
closest to the euclidean metric.  The gauge invariant state is then
chosen to be the restriction of the state to this gauge surface.  In
the spirit of \cite{florian}, we restrict to embedded spin networks
$S$ living on $\mathcal{T}$.  Given an $s$-knot $s$, there is only a
discrete number of such spin networks that are in the class $s$: we
choose $S(s)$ that maximizes (\ref{vacuum2}) among these.  We expect
this definition (possibly with an appropriate correction of the
Conrady vacuum state) to converge for fine triangulations, making the
background structure chosen effectively irrelevant for a triangulation
sufficiently finer than the Planck length.  This construction provides
a finite definition of $\Psi_{0}[s]$, diffeomorphism invariant by
definition.

(iv) Finally, we need to define the field $\phi(x)$ appearing in
(\ref{W4s}).  Following \cite{florian} we write $
h_{s}^{ab}(x) = ({q}_{S(s)}^{ab}(x) - \delta^{ab})$, where the point
$x$ is defined in terms of the boundary metric $q$ and 
${q}_{S(s)}^{ab}$ is
defined above (\ref{vacuum1}).  An alternative, which we do not pursue
here, is to derive $h^{ab}(x)(S)$ from the action of two $SU(2)$
generators \cite{simone}.

We can now bring together the various pieces discussed.  To start
with, consider a parallelepiped in 4d euclidean space with hight $T$
and cubic base of side $L$.  Let $\Sigma$ be its boundary, equipped
with the induced metric $q$.  Fix a triangulation of $\Sigma$.  The
simplest choice is to start from a cubic triangulation of $\Sigma$,
and to obtain a four-valent lattice, by splitting each (six-valent)
vertex of the cubic lattice into two vertices.  Replacing the various
items discussed into the formal expression (\ref{Wn4}) we obtain 
\begin{equation}
W^{a_{1}b_{1} \ldots a_{n}b_{n}}(x_{1}, \ldots x_{n}; L, 
T)=Z_{LT}^{-1}
 \sum_{s} c(s)\  {h}_{s}^{a_{1}b_{1}}({x_{1}})\ldots\,
{h}_{s}^{a_{n}b_{n}}({x_{n}})\ 
\Psi_{q}[s]\ 
 W[s] 
\label{W4D}
\end{equation}
where the sum is over all the $s$-knots that can be embedded in the
triangulation.  The normalization factor is the ``vacuum to vacuum"
amplitude $ Z_{LT}= \sum_{s} c(s)\ \Psi_{q}[s] \, W[s]$.  (\ref{W4D})
can be expanded in powers $\lambda^n$.  $n$ is the number of vertices
of the spinfoam, which is the number of 4-simplices in a simplicial
complex dual to the spinfoam, if this exists.  As a rough estimate, we
can imagine each 4-simplex to have Planck size: if classical
configurations dominate, the main contribution should come from $n$ of
the order of the 4-volume of the interaction region in Planck units.

Hypotheses and approximations used to get to (\ref{W4D}) are severe. 
But all quantities in (\ref{W4D}) are well defined.  The expression is
probably finite at any order in $\lambda$.  We can thus take
(\ref{W4D}) as a tentative definition of an $n$-point function within
the formalism of non-perturbative quantum gravity.  More precisely, we
can consider (\ref{W4D}) as a tentative concrete definition of the
quantity formally given by
\begin{eqnarray}
W^{a_{1}b_{1} \ldots a_{n}b_{n}}(x_{1}, \ldots, x_{n})= Z^{-1}\int 
Dg\ 
g^{a_{1}b_{1}}(x_1)
\ldots\, g^{a_{n}b_{n}}(x_{n})\ e^{-S_{EH}[g]}
\label{form}
\end{eqnarray}
where $S_{EH}$ is the Einstein-Hilbert action, computed at relative
spacetime distances $x_{1}, \ldots, x_{n}$ evaluated in terms of the
mean value of the quantum gravitational field itself, on a box
encircling the interaction region.

The construction can probably be ameliorated and varied in a number of
ways, and many issues remain open.  There are important missing steps
to get to the definition of quantities that can be interpreted as
particle transition amplitudes.  (On the physical interpretation of
``particle" states defined on finite spacial regions, see
\cite{daniele}.)  The key questions are whether the expression
(\ref{W4D}) is indeed finite, convergent, and independent from the
auxiliary structures uses to define it, when the triangulation is
sufficiently finer than the Planck scale, and whether this
construction leads, in a first approximation, to the general
relativity scattering tree amplitudes.

\end{document}